%% file: main.tex
\pdfoutput=1
\documentclass[a4paper]{article}

\usepackage{graphicx}
\usepackage{amsmath}
\usepackage{amssymb}
\usepackage{bm}
\usepackage{multirow}
\usepackage{booktabs}
\usepackage{subfigure}
\usepackage{mathptmx}
\usepackage{INTERSPEECH2020}

\title{Nonlinear Residual Echo Suppression Based on Multi-stream Conv-TasNet}
\name{Hongsheng Chen, Teng Xiang, Kai Chen, Jing Lu}
\address{
	Key Laboratory of Modern Acoustics, Institute of Acoustics, Nanjing University, Nanjing 210093, China.}
\email{hschen@smail.nju.edu.cn, txiang@smail.nju.edu.cn, chenkai@nju.edu.cn, lujing@nju.edu.cn}
\begin{document}
	
	\maketitle
	\begin{abstract}
		\input{abstract}
	\end{abstract}
	\section{Introduction}
	\input{introduction}
	\section{Model description}
	\input{model_description}
	\section{Simulations}
	\input{experimental}
	\section{Conclusion}
	\input{conclusion}
	\section{Acknowledgement}
	\input{acknowledgement}
	\bibliography{reference}
	
\end{document}

%% file: abstract.tex
Acoustic echo cannot be entirely removed by linear adaptive filters due to the nonlinear relationship between the echo and far-end signal. Usually a post processing module is required to further suppress the echo. In this paper, we propose a residual echo suppression method based on the modification of fully convolutional time-domain audio separation network (Conv-TasNet). Both the residual signal of the linear acoustic echo cancellation system, and the output of the adaptive filter are adopted to form multiple streams for the Conv-TasNet, resulting in more effective echo suppression while keeping a lower latency of the whole system. Simulation results validate the efficacy of the proposed method in both single-talk and double-talk situations.

%% file: introduction.tex
Acoustic echo is generated from the coupling between the loudspeaker and the microphone in full-duplex hands-free telecommunication systems or smart speakers. It will severely deteriorate the quality of speech communication and significantly degrade the performance of automatic speech recognition (ASR) within the smart speakers.\par

Typical linear acoustic echo cancellation (LAEC) methods use adaptive algorithms to identify the impulse response between the loudspeaker and the microphone \cite{hansler2005acoustic}. Frequency-domain least mean square algorithms are often utilized to guarantee both fast convergence speed and low computational load \cite{haykin2005adaptive}. The frequency-domain adaptive Kalman filter (FDKF) \cite{enzner2006frequency} is also a commonly used method with several efficient variations proposed recently \cite{yang2017frequency-domain,fan2019effective}.\par

The performance of LAEC methods severely degrades when nonlinear distortion is non-negligible in the acoustic echo path \cite{birkett1995limitations}. Usually a residual echo suppression (RES) module is required to further suppress the echo. The RES is usually conducted by estimating the spectral amplitude of the residual echo based on the far-end signal, filter coefficients and the residual signal of LAEC \cite{habets2008joint, valero2014signal-based, desiraju2020online, gustafsson1998combined, gustafsson2002a, chhetri2005regression}. However, it is difficult for the signal-processing-based RES to balance well between the residual echo attenuation and near-end speech distortion.\par

Recently, deep neutral network (DNN) has been introduced into RES due to its powerful capability of modeling nonlinear systems. The fully connected network (FCN) was employed to exploit multiple-input signals in RES \cite{carbajal2018multiple}. The drawback of the FCN is that it cannot effectively model the temporal structure of time series. The bidirectional long short-term memory (BLSTM) was also introduced to RES \cite{zhang2018deep}, but the non-casual processing increased the latency dramatically. The short time Fourier transform (STFT) is used to extract spectral amplitude features for these networks. However, high frequency resolution requires large signal blocks and thus leads to high latency for the calculation of STFT. Meanwhile, using the mask for spectral magnitude, as the training target cannot recover the phase of signal and limits the performance of the network \cite{wang2018supervised}.\par

RES can be regarded as a speech separation task focusing only on recovering the near-end signal from its mixture with the residual echo. The fully convolutional time-domain audio separation network (Conv-TasNet) \cite{luo2019conv} has been accepted as the state-of-the-art (SOTA) DNN-based solution in speech separation. Its performance is even better than that from the ideal time-frequency masks thanks to its pure end-to-end structure. In this paper, we modify the Conv-TasNet to effectively exploit the multiple streams created from the residual signal of LAEC and the output of the adaptive filter, and compare the performance of our proposed RES with several typical methods.\par


%% file: model_description.tex
\subsection{Problem formulation}
	\begin{figure}[htb]
		\centering
		\includegraphics[width=0.9\linewidth]{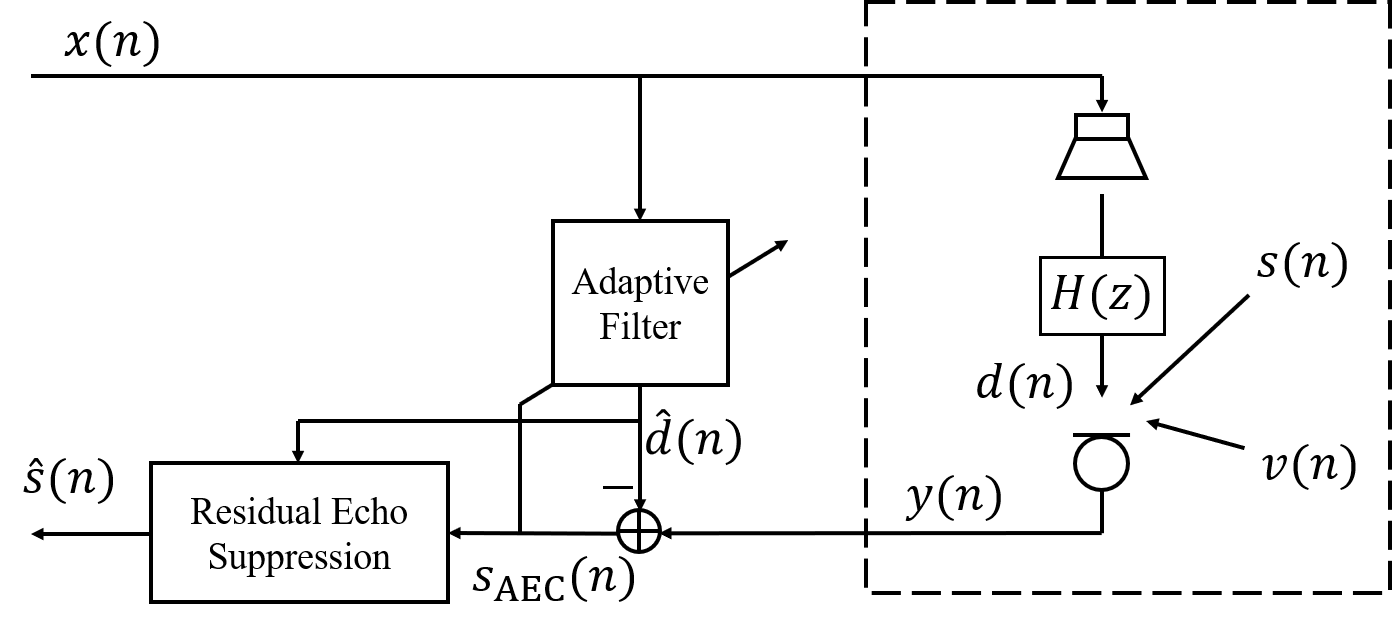}
		\caption[Diagram of AEC system with RES post-filter.]{Diagram of AEC system with RES post-filter.}
		\label{fig:aes}
	\end{figure}
	\noindent The AEC system with RES post-filter is depicted in Figure \ref{fig:aes}, where $ x(n) $ is the far-end signal, $ \hat{d}(n) $ is the output of the adaptive filter, $ H(z) $ represents the echo path transfer function, and the microphone signal $ y(n) $ consists of the echo $ d(n) $, the near-end speech $ s(n) $ and background noise $ v(n) $ as
	\begin{equation}
		y(n) = s(n) + d(n) + v(n).
	\end{equation}
	The signal of the LAEC $ s_{\rm AEC}(n) $ is given by subtracting the output of the adaptive filter $ \hat{d}(n) $ from the microphone signal $ y(n) $, with
	\begin{align}
		\hat{d}(n) = \hat{h}(n) * x(n)\\
		s_{\rm AEC}(n) = y(n) - \hat{d}(n)
	\end{align}
	where $ \hat{h}(n) $ denotes the adaptive filter and $ * $ represents convolution operation. Due to the inevitable nonlinear feature in the echo path, the LAEC cannot perfectly attenuate the echo, and $ s_{\rm AEC}(n) $ can be regarded as the mixture of the residual echo, background noise and the near-end signal. The RES can be designed from the viewpoint of speech separation, but unlike the standard speech separation, the auxiliary information can be extracted from the adaptive filter to improve the performance. In this paper, $ s_{\rm AEC}(n) $ and $ \hat{d}(n) $ are utilized to construct multiple streams for Conv-TasNet, the SOTA DNN-based solution for speech separation.

\subsection{Model design}
\begin{figure*}[tb!]
	\centering
	
	\subfigure[]{
		\begin{minipage}[t]{0.65\linewidth}
			\centering
			\includegraphics[height=0.31\textheight]{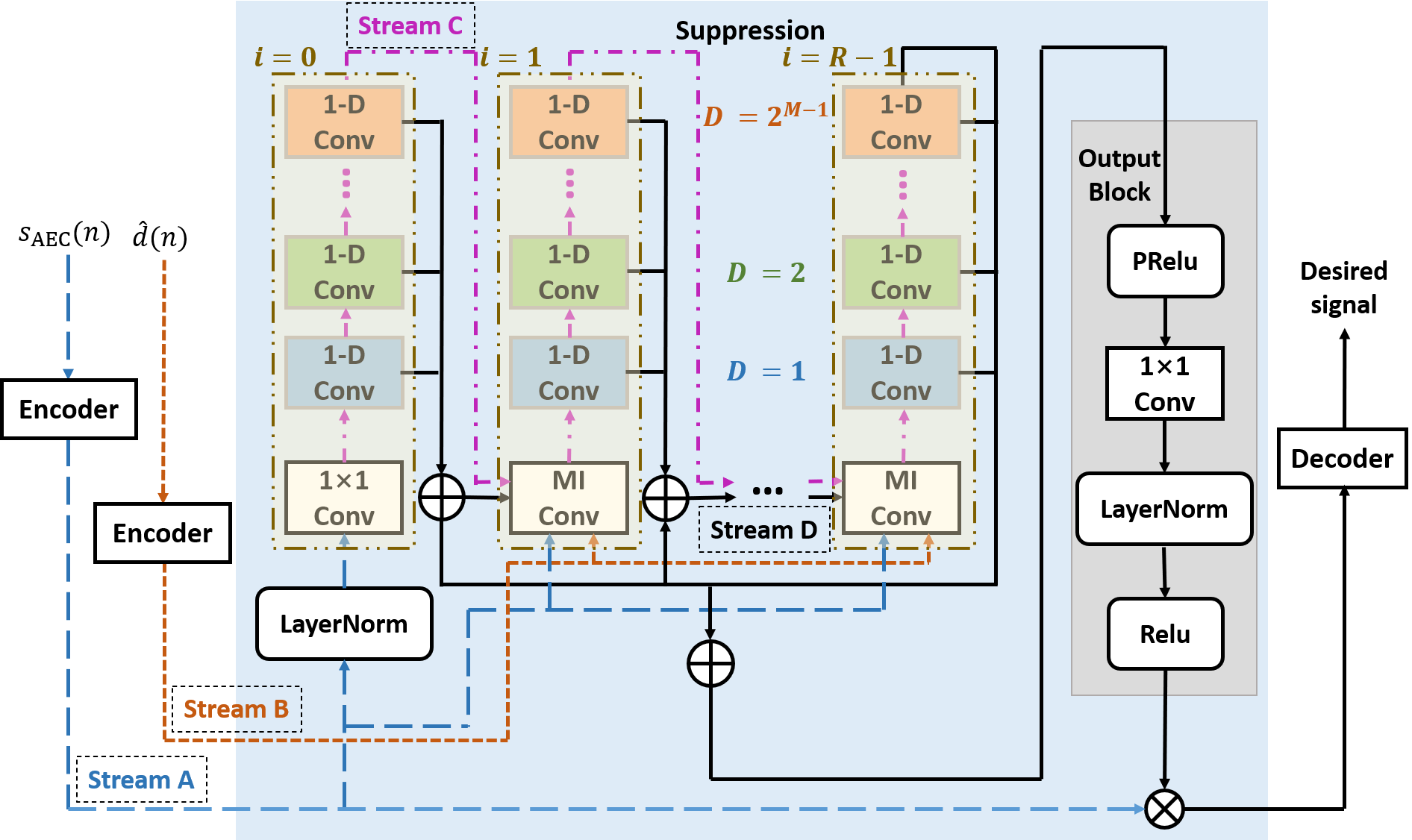}
		\end{minipage}%
	\label{fig:modifiedtasnet}
	}%
	\subfigure[]{
		\begin{minipage}[t]{0.35\linewidth}
			\centering
			\includegraphics[height=0.32\textheight]{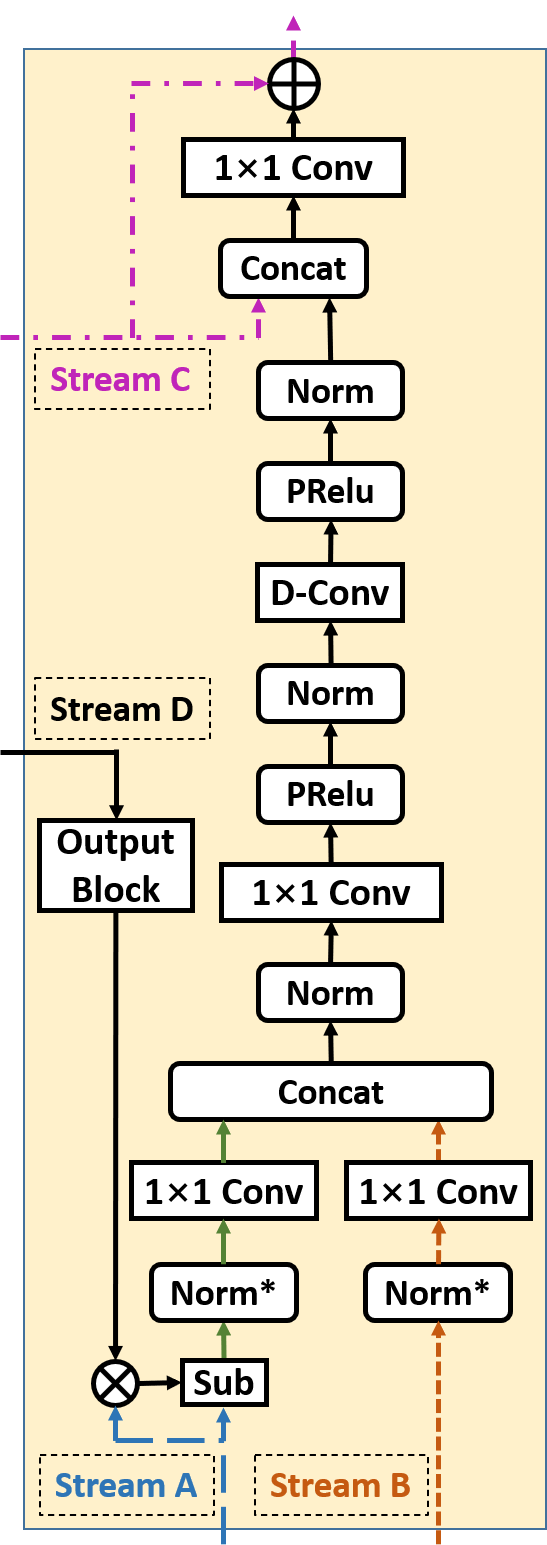}
		\end{minipage}%
	\label{fig:miconv}
	}%
	\centering
	\caption[The block diagram of the TasNet system.]{(a) The block diagram of the modified Conv-TasNet. (b) The block diagram of the MI Conv block. Steam A: the blue long-dash line, Steam B: the brown dash line, Steam C: the purple dash-dot line, Steam D: the black solid line.}
\end{figure*}

	Similar to the original Conv-TasNet, our model consists of encoder, suppression and decoder modules, as depicted in Figure \ref{fig:modifiedtasnet}. The encoder is a 1-D convolutional layer used to convert each frame of the waveform into effective representations and the decoder is a transposed convolutional layer used to invert the representations back to the waveform. The suppression module includes a temporal convolutional network (TCN), which is composed of $ R $ layers ($ i=0,1,…,R-1 $) and each layer contains $ M $ 1-D Conv blocks whose dilation factors are $ 1,2,4,...,2^{M-1} $ respectively. The output of the 1-D Conv block consists of two parts: the residual output (input of the next block) and the skip-connection path (part of the TCN’s output). The sum of the 1-D Conv blocks’ skip-connection path is then processed by the Output block to obtain the mask estimation. Every input to convolutional layers is zero padded to ensure the invariance of output length.\par
	
	From the LAEC processing procedure, it is straightforward to see that the output of the adaptive filter $ \hat{d}(n) $ is closely related to the residual echo. Therefore, we create two encoders for $ s_{\rm AEC}(n) $ and $ \hat{d}(n) $ respectively. In the suppression module, we add a multiple-input convolutional (MI Conv) block to each TCN’s layer to exploit the correlation between $ s_{\rm AEC}(n) $, $ \hat{d}(n) $ and the outputs of the shallow layers in TCN. The structure modification combined with the refined training objective (described in section 2.3) breaks the balance of channels in speech separation tasks and makes the network focus more on extracting the desired near-end signal. We also design an exponential layer normalization (eLN) operation to replace the cumulative layer normalization (cLN) in the original casual Conv-TasNet \cite{luo2019conv}, so that the system can work well when the variances of the near-end signal and the residual echo change rapidly over time.
	
	The exponential layer normalization operation is defined as:
	\begin{equation}
		eLN(\mathbf{f}_k) =\frac{\mathbf{f}_k-\hat{E}_k[\mathbf{f} ]}{(\hat{D}_k[\mathbf{f}] + \epsilon)^{\Omega}} \odot \gamma + \beta
	\end{equation}
	\begin{equation}
		\hat{E}_k[\mathbf{f}] = \dfrac{1-\alpha}{F} \sum_{p=0}^{N} {\left[\alpha^p \sum_{j=0}^{F-1} {\mathbf{f}_{k-p, j}}\right]}
		\label{eq:Ehat}
	\end{equation}
	\begin{equation}
		\hat{D}_k[\mathbf{f}] = \dfrac{1-\alpha}{F} \sum_{p=0}^{N} {\left[\alpha^p \sum_{j=0}^{F-1} \left(\mathbf{f}_{k-p, j}-\hat{E}_{k-i} [\mathbf{f}]\right)^2\right]}
		\label{eq:Dhat}
	\end{equation}
	where $ \odot $ denotes element-wise multiplication, $ \mathbf{f}_{k, j} \in \mathbb{R} $ is the $ j $-th feature of the $ k $-th frame, $ F $ is the feature dimension of $ \mathbf{f} $, $ \gamma,\beta \in \mathbb{R}^{F \times 1} $ are trainable parameters, and $ \alpha,\ \epsilon,\ \Omega $ are constants with $ \alpha $ the forgetting rate, $ \epsilon $ a robust regularization parameter and $ \Omega $ usually set to $ 0.5 $ unless otherwise specified. Equations (\ref{eq:Ehat}) and (\ref{eq:Dhat}) are exponential moving averages for estimating the means and variances of $ \mathbf{f}_k $ respectively if the parameter $ N $ is set to infinity. We choose a finite constant $ N $ instead of infinity for convenience of implementation in TensorFlow, which enables us to realize the eLN by conv operations.\par
	
	The diagram of the MI Conv block is shown in Figure \ref{fig:miconv}. The input of the MI Conv block consists of four streams: the representations of $ s_{\rm AEC}(n) $ (stream A) and $ \hat{d}(n) $ (stream B), the residual output of the previous block (stream C) and the sum of the previous layers’ skip-connection paths (stream D). Aiming at extracting the estimated residual echo feature from the output of previous layers, a Sub operation between stream A and stream D is utilized as:
	\begin{align}
		\mathbf{f_O}_i = g_{ob}(\mathbf{f_D}_{i-1}) \odot \mathbf{f_A}\\
		\mathbf{f_{sub}}_i = \mathbf{f_A} - \lambda \odot \mathbf{f_O}_i
	\end{align}
	where $ \mathbf{f_A}, \mathbf{f_D}_i $ are the features of Stream A and the $ i $-th Stream D, $ g_{ob} $ is the operation of the Output block (identical to the Output block in TCN), $ \mathbf{f_O}_i $ is the output of modules before the $ i $-th layers, $ \mathbf{f_{sub}}_i $ is the output of the Sub operation and $ \lambda \in \mathbb{R}_+^{F \times 1} $ is a trainable parameter. We regard $ \mathbf{f_O}_i $ as a proper approximation of the representations of the near-end signal, correspondingly, $ \mathbf{f_{sub}}_i $ represents the approximated information of the residual echo. We then partly normalize Stream B and the output of the Sub operation by Norm* layers and reduce their feature dimensions by $ 1 \times 1 $ Conv operations respectively. The Norm* layer is an eLN with $ \Omega $ set to $ 0.4 $. The outputs of the two $ 1 \times 1 $ Conv operation are concatenated together and then normalized. The following operations are $ 1 \times 1 $ Conv operation and depth-wise convolution (D-Conv) operation with parametric rectified linear unit (PRelu) \cite{he2015delving} followed by another normalization block. A $ 1 \times 1 $ Conv operation after concatenation serves as the residual path. All $ 1 \times 1 $ Conv operations in the MI Conv block have the same number of channels as the first $ 1 \times 1 $ Conv operation in the first layer of TCN, which is called the bottleneck layer. These operations aim to more effectively merge the features of $ \hat{d}(n) $ and the estimated residual echo into stream C.\par
	
\subsection{Training objective}
	The target function of training is the weighted sum of several loss functions, which is used to speed up training process and meet the requirement of the MI Conv block:
	\begin{equation}
		Loss_{total} = \frac{loss_{last} + \sum_{i=1}^{R-1} w^{R-i}loss_i}{1 + \sum_{i=1}^{R-1} {w^{R-i}}}
		\label{eq:losstotal}
	\end{equation}
	where $ loss_i $ represents the loss of the waveform of $ \mathbf{f_O}_i $ converted by the same decoder, $ loss_{last} $ is the loss of model output, and $ w \in [0, 1] $ is a weight coefficient. We choose scale-invariant source to noise ratio (SISNR) \cite{luo2019conv} for $ loss_i $ and $ loss_{last} $:
	\begin{align}
		\bm{s}_{target}&=\frac{\langle \hat{\bm{s}},\bm{s}\rangle \bm{s}}{\|\bm{s}\|^2 }\\
		\bm{e}_{noise}&=\hat{\bm{s}}-\bm{s}_{target}\\
		SISNR&=10\log_{10}{\frac{\|\bm{s}_{target}\|^2}{\|\bm{e}_{noise}\|^2}}
	\end{align}
	where $ \hat{\bm{s}}, \bm{s} $ are the estimated and original clean sources respectively, and $ \|\bm{s}\| $ denotes the $ l_2 \ norm $ of $ \bm{s} $.

%% file: experimental.tex
\subsection{Dataset}
Unlike telecommunication system, where the far-end signal is usually speech, music often acts as the `far-end' signal for smart loudspeakers. Therefore we use both speech and music as the far-end signal, and the near-end signal is speech. We use TIMIT \cite{lamel1989speech} as the speech database and MUSAN as the music database \cite{snyder2015musan}. We randomly choose 400 speakers for training and 40 different speakers for test. There are ten utterances for each speaker sampled at 16 kHz, one of which in the training set is spared for validation. The wave from MUSAN are split into 4-second segments, and 38577 segments are utilized for training, 400 segments for validation, and 400 segments for test. Totally we generate 74577 pieces of residual echo (36000 pieces of speech and 38577 pieces of music) for training, 800 pieces (400 pieces of speech and 400 pieces of music) for validation and 800 pieces (400 pieces of speech and 400 pieces of music) for test. Each epoch during training contains 3600 pairs of near-end signal and residual echo.\par

In order to generate simulated echo, a soft clipping, a sigmoidal function and a convolution operation are successively applied to the far-end signal. The soft clipping is defined as:
\begin{equation}
	Clip_{soft}(x(n)) = \dfrac{x_{max}x(n)}{\sqrt{|x_{max} |^2+|x(n)|^2}}
\end{equation}
where $ x_{max} $ determines the maximum value of the soft clipping, which is set to $ 80\% $ of the maximum value of the input signal. The sigmoidal function \cite{comminiello2017full} is an approximation to the non-linearity of a loudspeaker:
\begin{equation}
	NL\left(x\left(n\right)\right) = \frac{1}{1+e^{\left[-a\cdot b\left(n\right)\right]}}-\frac{1}{2}
\end{equation}
\begin{equation}
	b\left(n\right) = \frac{3}{2}x\left(n\right)-\frac{3}{10}x^2\left(n\right)
\end{equation}
\begin{equation}
	a = 
	\begin{cases}
	4 & \text{if } b(n) > 0\\
	2 & \text{if } b(n) \le 0
	\end{cases}
\end{equation}
The value of $ a $ is set to 2 instead of 0.5 when $ b(n) \le 0 $ for smoother non-linearity. For the convolution operation, we construct 50 simulated rooms, each dimension of which is randomly chosen from $ [2, 5]\ {\rm m}$ and $ T_{60} $ is randomly chosen from $ [150, 450]\ {\rm ms} $. A microphone and a loudspeaker are randomly placed in each room for generation of room impulse responses (RIRs) by the image method \cite{lehmann2009diffuse}. Four hundred of these RIRs are used to generate the training set while the rest one hundred RIRs are used to generate the test set.\par

The original signal to echo ratio (SER) (before processing of LAEC) is randomly chosen from $ \{-12.2, -14.2, -16.2, -18.2\}\ {\rm dB} $ and the Gaussian white noise is added with signal to noise ratio (SNR) randomly chosen from $ \{30, 20\}\ {\rm dB} $. The frequency-domain adaptive Kalman filter \cite{enzner2006frequency} acts as the LAEC, and the quartiles of its echo attenuation in training set is about $ 13.3 \ {\rm dB} $, $ 14.0 \ {\rm dB} $ and $ 14.8 \ {\rm dB} $ respectively.\par


\subsection{Simulation configurations}
The total latency of network is set to 240 samples (15 ms). The number of filters in the encoder is 512, and the length of each filter is 40 with $ 75\% $ overlap. The number of channels in the bottleneck layer, the residual paths and the skip-connection paths is 256. The numbers of channels and the kernel size in the 1-D Conv blocks are 512 and 3, respectively. The kernel size in the MI Conv blocks' D-Conv layers is 128. The number of repeats $ R $ is 4 and each repeat layer has 8 1-D Conv blocks.
Hyperparameters $ \alpha $, $ N $ and $ w $ are set to $ \sqrt[640]{0.001} $, $ 640 $ and $ \sqrt{1/2} $ respectively.\par

The model is trained by Adam optimizer \cite{kingma2014adam} for 120 epochs with each batch containing two pairs of 4 s signals sampled at 16 kHz. The initial learning rate is set to 1e-3 and is halved every time the validation loss is not improved in four continuous epochs. We apply $ l_2 \ norm $ gradient clipping with a maximum of 5. TensorFlow is employed for model implementation and two Nvidia GeForce GTX 1080Ti are used for training.\par


In the following simulations, we name our proposed method as TasNet-MI and compare its performance with four other models: FCN \cite{carbajal2018multiple}, BLSTM \cite{zhang2018deep}, the original Conv-TasNet (TasNet-O) \cite{luo2019conv} and the modified Conv-TasNet without the information of $ \hat{d}(n) $ (TasNet-L). The TasNet-O uses the same overlap and normalization method as the TasNet-MI. Considering that the MI Conv blocks increase the network’s capacity, we add a 1-D Conv block to each layer (except the first layer) of the TasNet-O’s TCN. Compared with TasNet-MI, TasNet-L only neglects the stream of $ \hat{d}(n) $, which is more appropriate to verify the advantage of exploiting information of $ \hat{d}(n) $. Exemplary audio samples are available online at ``https://github.com/Mo-yun/tasnetmi-samples''.

\subsection{Evaluation metrics}
In double-talk situations, we use three metrics for performance evaluation: perceptual evaluation of speech quality (PESQ) \cite{rix2001perceptual}, signal to distortion Ratios (SDR) \cite{vincent2006performance, raffel2014mir_eval} and short-Time objective intelligibility (STOI) \cite{taal2010short}. We also use echo return loss enhancement (ERLE) as metrics for echo attenuation in single-talk situations \cite{enzner2014acoustic}.

\subsection{Results}
The extra ERLE over LAEC at single-talk situations for both speech and music echo are presented in Table \ref{tab:ERLE}. It can be seen that BLSTM has the best performance. The TasNet-MI achieves over 40 dB ERLE for both speech and music, significantly better than the other two TasNet models, illustrating the benefit of the multi-stream information. The ERLE of the FCN is limited since it cannot fully utilize relations between time series. However, it should be noted that over 15 dB extra ERLE in single-talk situations is sufficient for most practical applications.\par
\begin{table}[htb]
	\caption{Average ERLE in single-talk situation}
	\label{tab:ERLE}
	\centering
	\begin{tabular}{llcc}
		\toprule
		Far-end signal & & speech & music\\
		\midrule
		\multirow{5}{*}{ERLE} & FCN & 16.55 & 18.93 \\
		& BLSTM & 51.67 & 55.81 \\
		& TasNet-O & 29.14 & 32.30 \\
		& TasNet-L & 35.27 & 41.70 \\
		& TasNet-MI & 45.33 & 47.77 \\
		\bottomrule
	\end{tabular}
\end{table}
The double-talk scenario is more challenging due to the difficulty of balancing between the residual echo suppression and the quality of the near-end speech. In our test, the Gaussian white noise is added to near-end signal with 30 dB SNR, and the far-end signal is speech or music with the original SERs (before processing of the LAEC) set to $ -14.2\ {\rm dB} $ and $ -18.2\ {\rm dB} $. The performance in terms of PESQ, SDR and STOI is shown in Tables \ref{tab:speech} and \ref{tab:music}. The performance of the TasNet-O is better than the FCN and the BLSTM in almost all these conditions, showing the benefit of the end-to-end time-domain solution, which has already been validated in speech separation. The improvement of the TasNet-L over the TasNet-O validates the modification of training target and network structure, which aims to make the system focus more on extracting the desired near-end signal. Moreover, the TasNet-MI further outperforms the TasNet-L, validating that the MI Conv block provides an effective way to exploit the information of $ \hat{d}(n) $, leading to more residual echo suppression while recovering the near-end signal with higher quality.

\begin{table}[htb]
	\caption{Average PESQ, SDR and STOI when the far-end signal is speech in double-talk situation}
	\label{tab:speech}
	\centering
	\begin{tabular}{llcc}
		\toprule
		SER & & $ -14.2\ {\rm dB} $ & $ -18.2\ {\rm dB} $\\
		\midrule
		\multirow{6}{*}{PESQ} & LAEC Only & 1.62 & 1.36 \\
		& FCN & 2.25 & 1.89 \\
		& BLSTM & 2.45 & 2.16 \\
		& TasNet-O & 2.57 & 2.25 \\
		& TasNet-L & 2.71 & 2.41 \\
		& TasNet-MI & 2.80 & 2.50 \\
		\hline
		\multirow{6}{*}{SDR}  & LAEC Only & $ -0.99 $ & $ -4.88 $ \\
		& FCN & 8.06 & 4.73 \\
		& BLSTM & 8.04 & 5.10 \\
		& TasNet-O & 11.7 & 9.10 \\
		& TasNet-L & 12.6 & 10.1 \\
		& TasNet-MI & 13.8 & 11.3 \\
		\hline
		\multirow{6}{*}{STOI} & LAEC Only & 0.626 & 0.531 \\
		& FCN & 0.784 & 0.695 \\
		& BLSTM & 0.851 & 0.791 \\
		& TasNet-O & 0.879 & 0.813 \\
		& TasNet-L & 0.895 & 0.835 \\
		& TasNet-MI & 0.912 & 0.860 \\
		\bottomrule
	\end{tabular}
\end{table}

\begin{table}[htb!]
	\caption{Average PESQ, SDR and STOI when the far-end signal is music in double-talk situation}
	\label{tab:music}
	\centering
	\begin{tabular}{llcc}
		\toprule
		SER & & $ -14.2\ {\rm dB} $ & $ -18.2\ {\rm dB} $\\
		\midrule
		\multirow{6}{*}{PESQ} & LAEC Only & 1.51 & 1.26 \\
		& FCN & 2.13 & 1.77 \\
		& BLSTM & 2.32 & 2.01 \\
		& TasNet-O & 2.38 & 1.97 \\
		& TasNet-L & 2.53 & 2.15 \\
		& TasNet-MI & 2.59 & 2.19 \\
		\hline
		\multirow{6}{*}{SDR}  & LAEC Only & $ -2.42 $ & $ -6.31 $ \\
		& FCN & 7.66 & 4.23 \\
		& BLSTM & 7.60 & 4.66 \\
		& TasNet-O & 11.6 & 8.67 \\
		& TasNet-L & 12.4 & 9.60 \\
		& TasNet-MI & 13.1 & 10.4 \\
		\hline
		\multirow{6}{*}{STOI} & LAEC Only & 0.625 & 0.536 \\
		& FCN & 0.751 & 0.652 \\
		& BLSTM & 0.828 & 0.763 \\
		& TasNet-O & 0.862 & 0.773 \\
		& TasNet-L & 0.879 & 0.800 \\
		& TasNet-MI & 0.894 & 0.820 \\
		\bottomrule
	\end{tabular}
\end{table}

%% file: conclusion.tex
An effective DNN-based residual echo suppression is proposed in this paper based on the modification of the Conv-TasNet. We adopt the residual signal of linear acoustic echo cancellation system and the output of the adaptive filter to form multiple streams, and utilize the extra multi-input Conv blocks to effectively merge the information of the streams into the network, aiming at suppressing more residual echo and recovering high-quality near-end speech. Simulation results validate the efficacy of the proposed method in both single-talk and double-talk situations.

%% file: acknowledgement.tex
The National Science Foundation of China supported this work with grant number 11874219.